\documentclass[traditabstract]{aa} % for the abstract without structuration
\usepackage{graphicx}
\usepackage{txfonts}
\usepackage{booktabs}
\usepackage{multirow}
\usepackage{natbib}
\usepackage{wasysym} 
\usepackage{wrapfig,lipsum,booktabs}

\bibpunct{(}{)}{;}{a}{}{,}
\newcommand{\teff}{\mbox{$\rm T_{\rm eff}$}}
\newcommand{\tprim}{\mbox{$\rm T_{\rm eff,1}$}}
\newcommand{\tsec}{\mbox{$\rm T_{\rm eff,2}$}}
\newcommand{\logg}{\mbox{$\log g_1$}}
\newcommand{\loggs}{\mbox{$\log g_2$}}
\newcommand{\vsini}{\mbox{$v \sin i$}}

\newcommand{\kms}{\mbox{km\,s$^{-1}$}}

\newcommand{\rsun}{R\ensuremath{_\odot}}
\newcommand{\msun}{M\ensuremath{_\odot}}
\newcommand{\rprim}{R\ensuremath{_1}}
\newcommand{\rsec}{R\ensuremath{_2}}

\newcommand{\mstar}{\ensuremath{M_\star}}
\newcommand{\porb}{\ensuremath{P_{\rm orb}}}
\newcommand{\asini}{\ensuremath{a \sin{i}}}
\newcommand{\potp}{\ensuremath{\Omega_1}}
\newcommand{\pots}{\ensuremath{\Omega_2}}
\newcommand{\eblm}{J0113+31}

\def\secos{$\sqrt{e} \cos \omega$}
\def\sesin{$\sqrt{e} \sin \omega$}
\def\feh{[Fe/H]}

\begin{document}

   \title{EBLM Project II }

  \subtitle{
A very hot, low-mass M dwarf in an eccentric and long period \\ eclipsing binary system from SuperWASP
}

  \author{Y. G\'omez Maqueo Chew \inst{\ref{warwick}}   \and 
	  J. C. Morales \inst{\ref{ieec},\ref{lesia}} \and 
          F. Faedi \inst{\ref{warwick}}  \and 
	  E. Garc\'ia-Melendo \inst{\ref{upv},\ref{edo},\ref{ieec}} 
\and
	  L. Hebb \inst{\ref{hws}} \and F. Rodler \inst{\ref{cfa}}  \and R. Deshpande \inst{\ref{cehw},\ref{psu}} 
	  \and S.~Mahadevan \inst{\ref{cehw},\ref{psu}}  
\and J. McCormac  \inst{\ref{ing},\ref{warwick}} 
\and R.~Barnes \inst{\ref{uw}} 
\and A.~H.~M.~J.~Triaud \inst{\ref{mit},\ref{fellow}}
	\and M. Lopez-Morales \inst{\ref{ieec},\ref{cfa}} 
	\and I. Skillen \inst{\ref{ing}} 
	  \and A. Collier Cameron \inst{\ref{sta}} \and  M. D. Joner \inst{\ref{utah}} \and C. D. Laney  \inst{\ref{utah}} \and D. C. Stephens \inst{\ref{utah}}
\and K.~G. Stassun \inst{\ref{vandy},\ref{fisk}}
\and P. Monta\~n\'es-Rodr\'iguez \inst{\ref{iac},\ref{laguna}}
}

\institute{Department of Physics, University of Warwick, Coventry CV4 7AL, UK \label{warwick}\\
  \email{y.gomez@warwick.ac.uk}
  \and Institut d'Estudis Espacials de Catalunya (IEEC), Edif. Nexus, C/ Gran Capit\`a 2-4, 08034 Barcelona, Spain \label{ieec}
  \and LESIA-Observatoire de Paris, CNRS, UPMC Univ. Paris 06, Univ. Paris-Diderot, France \label{lesia}
\and Departamento de F\'isica Aplicada I, E.T.S. Ingenier\'ia, Universidad del Pa\'is Vasco, 9 Alameda Urquijo s/n, 48013 Bilbao, Spain \label{upv}
\and Esteve Duran Observatory Foundation Fundaci\'o Observatori Esteve Duran. Avda. Montseny 46, Seva 08553, Spain \label{edo}
\and Department of Physics, Hobart and William Smith Colleges, Geneva, New York, 14456, USA \label{hws} 
 \and Harvard-Smithsonian Center for Astrophysics, 60 Garden St, Cambridge, MA 02138, USA \label{cfa} 
\and Center for Exoplanets and Habitable Worlds, The Pennsylvania State University, University Park, PA 16802, USA \label{cehw}
  \and Department of Astronomy and Astrophysics, The Pennsylvania State University, University Park, PA 16802, USA \label{psu}
   \and Isaac Newton Group of Telescopes, Apartado de Correos 321, E-38700 Santa Cruz de Palma, Spain \label{ing}
  \and Astronomy Department, University of Washington, Seattle, Washington, USA \label{uw}
  \and Kavli Institute for Astrophysics \& Space Research, Massachusetts Institute of Technology, Cambridge, MA 02139, USA \label{mit}
\and  Fellow of the Swiss National Science Foundation \label{fellow}
  \and School of Physics and Astronomy,  University of St. Andrews, St. Andrews, Fife KY16 9SS, UK \label{sta} 
  \and Department of Physics and Astronomy, N283 ESC, Brigham Young University, Provo, UT 84602-4360, USA \label{utah} 
\and Physics and Astronomy Department, Vanderbilt University, Nashville, Tennessee, USA \label{vandy}
\and Department of Physics, Fisk University, Nashville, TN 37208 USA \label{fisk}
\and Instituto de Astrof\'isica de Canarias, C/V\'ia L\'actea sn, 382000, La Laguna, Tenerife , Spain \label{iac}
\and  Departamento de Astrof\'isica, Universidad de La Laguna, Av., Astrof\'isico Francisco S\'anchez, sn, E38206, La Laguna, Spain \label{laguna}
   }

   \date{Received ; accepted 28 August 2014}

% \abstract{}{}{}{}{} 
% 5 {} token are mandatory
 
   \abstract { 

In this paper, we derive the fundamental properties of 1SWASPJ011351.29+314909.7 (\eblm), 
a metal-poor ($-0.40$ $\pm$ 0.04 dex), eclipsing binary in an eccentric orbit ($\sim$0.3) with an orbital period of $\sim$14.277 d.  
Eclipsing M~dwarfs orbiting  
solar-type stars (EBLMs), like \eblm, have been identified from WASP light curves 
and follow-up spectroscopy in the course of the transiting planet search. 
We present the first binary of the EBLM sample to be fully analysed, 
and thus, define here the methodology.  
The primary component with a mass of 0.945 $\pm$ 0.045 \msun\ has a large radius (1.378 $\pm$ 0.058 \rsun) indicating 
that the system is quite old, $\sim$9.5 Gyr. 
The M-dwarf secondary mass of 0.186 $\pm$ 0.010 \msun\ and radius of 
0.209 $\pm$ 0.011 \rsun\ are fully consistent with stellar evolutionary models. 
However, from the near-infrared secondary eclipse light curve, 
 the M~dwarf is found to have an effective temperature of 3922 $\pm$ 42~K, which is $\sim$600~K hotter than predicted by theoretical models.  We discuss different scenarios to explain this temperature discrepancy.  
The case of \eblm\  for which we can measure mass, radius, temperature and metallicity,
 highlights the importance of deriving mass, radius {\it and} 
temperature as a function of metallicity for M~dwarfs to better understand the lowest mass stars.
The EBLM Project will define the 
relationship between mass, radius, temperature and metallicity for M~dwarfs
providing important empirical constraints at the bottom of the main sequence.
}


   \maketitle
%
%________________________________________________________________

\section{Introduction}

The primary goal of NASA's forthcoming exoplanet mission, Transiting Exoplanet Survey Satellite (TESS), is to detect small 
transiting planets around bright, nearby host stars.  Due to the increased signal-to-noise in
the spectroscopic observations obtained for a bright host star, it is much easier to derive
both the mass of an orbiting planet using the radial velocity technique and 
to measure the spectroscopic signatures
from the planet's atmosphere.
In order to identify bright planet hosting stars over the whole sky,  
TESS will reside in a High-Earth Orbit and continuously monitor each field of target stars
for 27 consecutive days.  While stars that reside in regions where the fields overlap will 
have longer duration light curves, the majority of newly discovered TESS planets
will have relatively short orbital periods ($< 27$~days).  
The consequence of this observing strategy is that potentially habitable worlds where liquid water could
exist on the surface will only be found around cool, very late type M~dwarf stars.  Luckily, intrinsically faint M~dwarf stars also have 
relatively high contrast ratios between the star and the orbiting planet, thus these systems will be 
optimal targets for detecting atmospheric signatures from the planet itself.  

As always, in order to fully understand the planets, it is paramount to accurately characterise their host stars.
The mass of a planet hosting star which directly determines the derived planet mass is typically obtained by comparing measurable star
properties (e.g., colours, \teff, luminosity) to theoretical stellar evolution models \citep[e.g.,][]{Dotter2008} and/or empirical relationships \citep[e.g.,][]{Torres2010}.
Although M~dwarfs comprise the majority of stars in the Galaxy, our understanding of the relationships between
their masses, radii, temperatures, and metallicities is still incomplete---particularly at the very bottom of the main sequence.
For example, recent analysis of a newly discovered $\sim 0.2$~M$_{\odot}$ eclipsing M~dwarf (KIC~1571511)
suggests its temperature is hotter than theoretical models predict by $\sim 900$~K \citep{Ofir2012}. 
If this is typical despite the accurate radii benefiting from Gaia parallaxes, the mass of the planets identified by TESS that are orbiting stars in the habitable zone will 
not be accurately characterised. 
A temperature that is hotter than expected for planet hosting M~dwarfs would mistakenly imply a higher stellar mass and consequently a higher mass for the planet.  
This erroneous characterisation is not unique to the transiting method, but would also affect other kinds of systems, for example those discovered astrometrically by Gaia \citep{Perryman2001}.

For low mass stars, the majority of existing knowledge necessary to calibrate
relationships between fundamental properties has come from two types of systems:
(1) nearby, single stars with interferometric radii measurements \citep[][and references therein]{Segransan2003,Berger2006,Demory2009,boyajian2012} 
and (2) M+M~dwarf eclipsing binaries \citep[EBs;][]{Lacy1977,Leung1978, Metcalfe1996,Torres2002,
Delfosse1999,Ribas2003,Lopez-Morales2005, Morales2009, Nefs2013}.  
However, the number of very late type stars (\mstar $\le$0.25\msun)
for which these analyses can be done is extremely small.
In the literature, there are only 18 measurements of stellar mass and radius of very low mass stars; of those, only 7 have temperatures
\citep[][and references therein]{Nefs2013,Zhou2014}.  Furthermore,
deficiencies are inherent in these techniques that contribute to our insufficient
knowledge of the properties of very late-type M~dwarfs.
Specifically, single interferometric systems allow for the measurement of radius, effective temperature (\teff), 
and sometimes metallicity (with large uncertainties), but not mass.  Furthermore,
M+M EBs provide direct measurements of the masses and radii of both components and
their relative temperatures, but individual temperatures and metallicities are difficult to derive from
the complex spectra of two unresolved M~dwarfs.  

To address these issues, we have an ongoing program designed to measure significant 
numbers of masses and radii of very low mass stars
with accurate metallicity and temperature determinations by analyzing
M~dwarfs in eclipsing systems with higher mass F, G, or K~stars \citep{Triaud2013}.
Hereafter, we refer to these systems as EBLMs.  
In an EBLM, the primary star dominates the light allowing an 
accurate temperature and metallicity to be determined from the relatively 
simple and well-understood spectrum of the F/G/K star.   
The mass of the primary is derived
from its stellar parameters and then used in combination with the radial velocity curve
and light curve to find the mass of the M~dwarf and the radius of both components, as well as the M~dwarf temperature.
Finally, the metallicity derived for the primary star is adopted for the M~dwarf secondary by assuming
the close binary pair formed from the same parent molecular cloud.

Our program is based around a large sample of EBLM systems ($\sim150$) discovered in the SuperWASP survey.
SuperWASP \citep{Pollacco2006} is a dedicated, ultra-wide field sky
survey, continuously monitoring stars of V$\sim$ 8--15 mag over a
quarter of the sky every clear night. The SuperWASP archive is a rich
source of new eclipsing binaries and in particular systems with low
mass secondaries are found in the course of exoplanet candidate
selections. These objects are bright and their light curves have
thousands of data points obtained over multiple years showing clear
eclipse signatures with well defined periods. Although EBLMs
are sources of false alarm detections when searching for exoplanets,
they are ideal objects to use for determining fundamental parameters
of very low mass stars.  

Here we present the study and characterisation of a newly discovered
EBLM from the SuperWASP survey, \eblm. The binary is
composed of a G0--G2 V primary and an
M$_{2}= 0.19 $M$_{\odot}$ secondary. The system is eccentric with
an orbital period of 14.277 days.  Like KIC~1571511, we find the temperature
of the M~dwarf to be significantly hotter than stellar evolution
models predict for a star of this mass and metallicity.
The following paper describes our analysis of this interesting system.
It is structured as follows: in \S \ref{obs} we describe the photometric
and spectroscopic observations utilised in the analysis of the eclipsing binary;
 in \S \ref{analysis}, we
describe our data analysis and our approach
to the modelling of the system, and in \S \ref{disc}, we discuss our results
and set them in context of other M-dwarf measurements.
Finally, we draw our conclusions in \S 5.

%__________________________________________________________________

\section{Observations and Data Reduction}\label{obs}

In this section, we describe the data utilised to derive the physical properties of the
orbit and the stellar components of the eclipsing binary. 
Although \eblm\ was discovered from its SuperWASP time-series
photometry, the survey-quality light curve  (\S \ref{wasp}; Fig.~\ref{waspLC})
does not have sufficient precision to model the eclipses.  
Thus, we obtained higher precision light curves of the primary eclipse 
at optical wavelengths, with the 0.4-m Near Infra-red Transiting Exoplanet Survey (NITES) telescope at La Palma (\S \ref{nites}), the 0.6-m telescope at the Observatori Esteve Duran (OED; \S \ref{oed}),  and the 0.91-m telescope at the BYU West Mountain Observatory
(\S \ref{byu}).
We also acquired a light curve of the secondary eclipse in the near-infrared 
 2.1-m telescope using the instrument FLAMINGOS at the Kitt Peak National Observatory (KPNO; \S \ref{sececl}).  
Additionally, we characterised the reflex motion of the primary star around the system's 
centre of mass via radial velocity measurements (\S \ref{not}--\ref{het}).

\subsection{Photometry--Optical: SuperWASP}\label{wasp}

\begin{figure}
\centering
\includegraphics[width=0.49\textwidth]{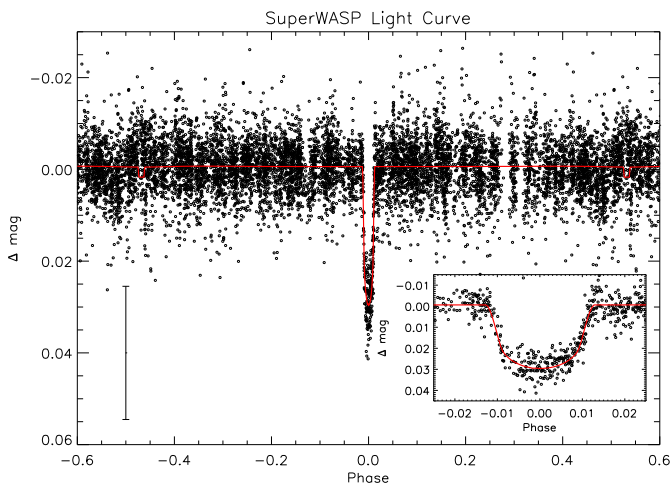}
\caption{SuperWASP Light Curve. 
We show the time-series photometry that led to the
discovery of \eblm, as part of the SuperWASP 
transiting planet survey, overplotted is the final model 
light curve from the EB modelling (\S \ref{eb}) in the $V$-band, used as a proxy for the 
non-standard SuperWASP filter.
The primary eclipse of the M dwarf blocking the light of the more massive star 
has a transit-like shape (shown in inset). 
The error bar to the bottom left represents the median uncertainty of the 
SuperWASP photometric points,
which are excluded for clarity. 
}
\label{waspLC}
\end{figure}

The SuperWASP North telescope is located in La Palma (ORM -
Canary Islands). 
The telescope consists of 8 Canon 200mm f/1.8 focal lenses coupled to
e2v $2048\times2048$ pixel CCDs, which yield a field of view of
$7.8\times7.8$ square degrees, and a pixel scale of 13.7\arcsec\
\citep{Pollacco2006}.  

\eblm\ was observed from 2004 June 11 through to 2011 January 16, 
with a total of 7 cameras and 29324 photometric points covering 357
primary eclipses with a total of 5607 points during eclipse. 
The SuperWASP data were first processed with the
custom-built reduction pipeline described in Pollacco et
al. (2006). The resulting light curves were analysed using our
implementation of the Box Least-Squares fitting an SysRem
de-trending algorithms (see \citealt{Cameron2006,Kovacs2002,Tamuz2005}), 
to search for signatures of planetary
transits. Once the candidate planet is flagged, a series of
multi-season, multi-camera analyses are performed to strengthen the
candidate detection. In addition different de-trending algorithms
(e.g., TFA, \citealt{Kovacs2005}) were used on one season and
multi-season light curves to confirm the transit-like signal and the
physical parameters of the putative planet candidate. These additional tests
allow a more thorough analysis of the system
parameter derived solely from the SuperWASP data thus helping in the
identification of the best candidates, as well as to reject possible
spurious detections. 
At this time \eblm\ was flagged as a candidate
exoplanet. Subsequently, during eye-balling of the SuperWASP light curves,
their morphology and the characteristic of the photometric signal lead to revise \eblm\ as an
EBLM. The SuperWASP light curve folded on the ephemerides derived in section \S\ref{mcmc} 
is shown in Figure \ref{waspLC}.

\subsection{Photometry--Optical: NITES}\label{nites}
The NITES telescope is an f/10 0.4-m Meade LX200GPS with advanced coma-free optics. The CCD camera is a Finger 
Lakes Instrumentation (FLI) Proline 4710
with a back illuminated 1024$\times$1024, 13$\mu$m/pixel, deep-depleted CCD made by e2v (for more details see \citealt{McCormac2014}).

Two partial light curves were observed with the NITES telescope
on La Palma on the nights of 2011 September 18, and 2012 December 18. 
The telescope was defocused to
5\arcsec~and 2385 images of 5s exposure time were obtained during the September observations. On
the December observations, the telescope was defocused to 4.6\arcsec~and 1105 images
with 5s exposure time were obtained, data on both nights was taken
without a filter. The data were bias subtracted, corrected for dark
current and flat fielded using the standard routines in IRAF and
aperture photometry was performed using DAOPHOT \citep{Stetson1987}. 
The stars GSC~2295-0229 and  GSC N322212139 (01:13:35.3 +31:49:01 J2000) 
where used as comparison stars. 
Other stars in the field were rejected to minimise the scatter in the light curve.
Shown in the bottom panel of Fig.~\ref{figlcs}, this data was only used in \S \ref{mcmc} to derive the best 
ephemeris but was excluded from the EB analysis from which the properties of the M dwarf are derived (\S \ref{eb}).

\subsection{Photometry--Optical: OED}\label{oed}

\begin{figure}[!ht]
\centering
\includegraphics[width=0.49\textwidth]{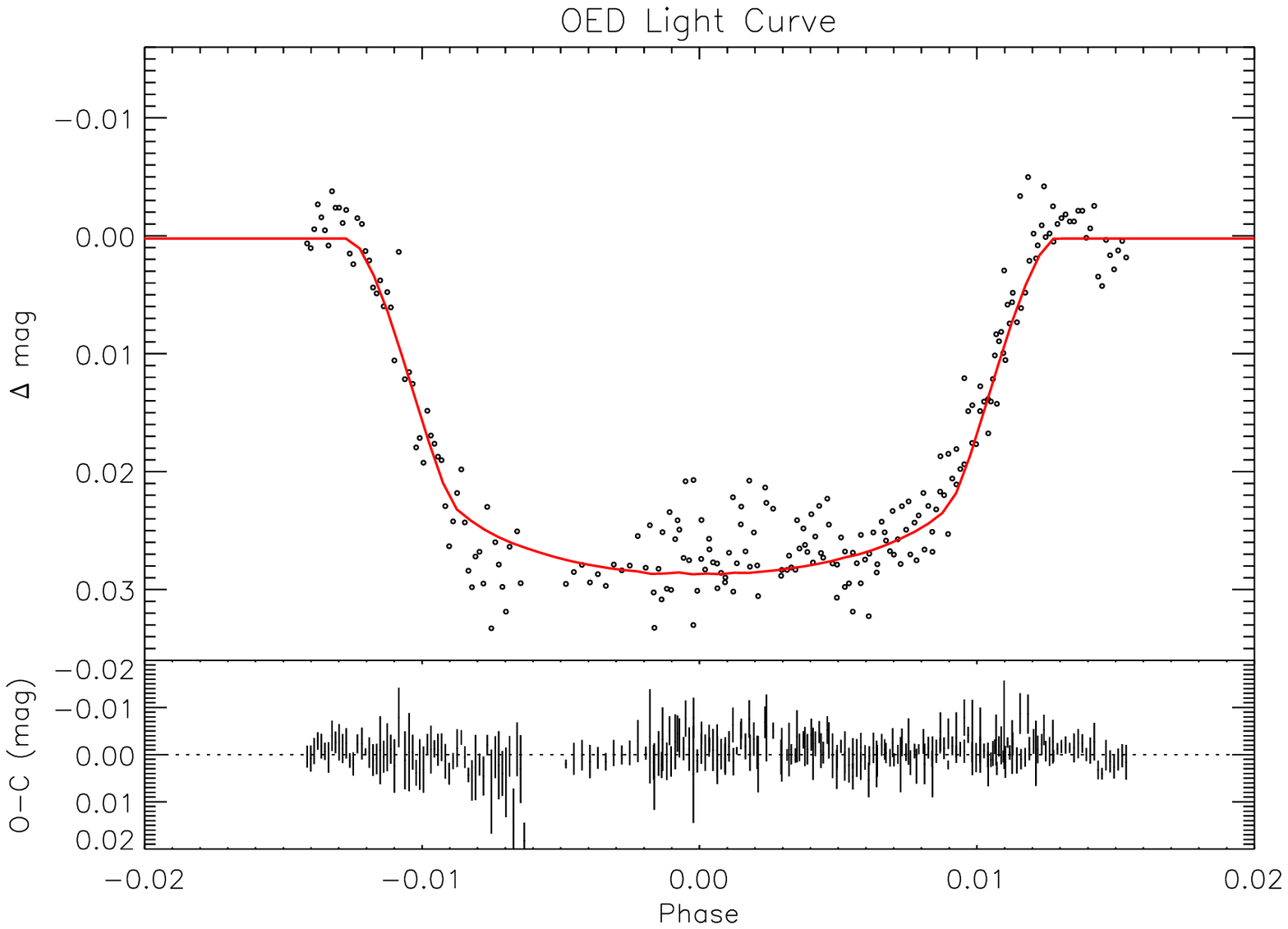}
\includegraphics[width=0.49\textwidth]{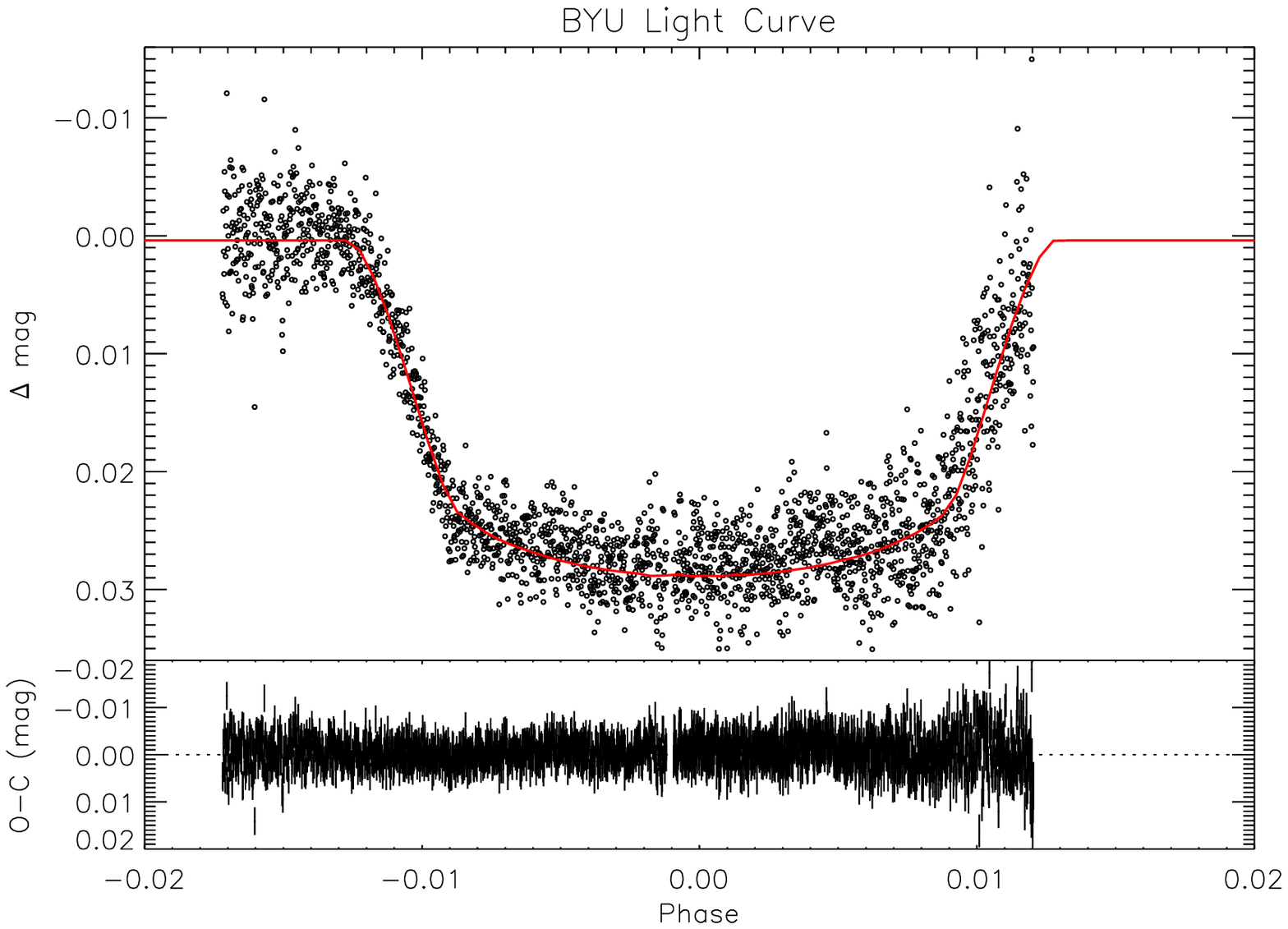}
\includegraphics[width=0.49\textwidth]{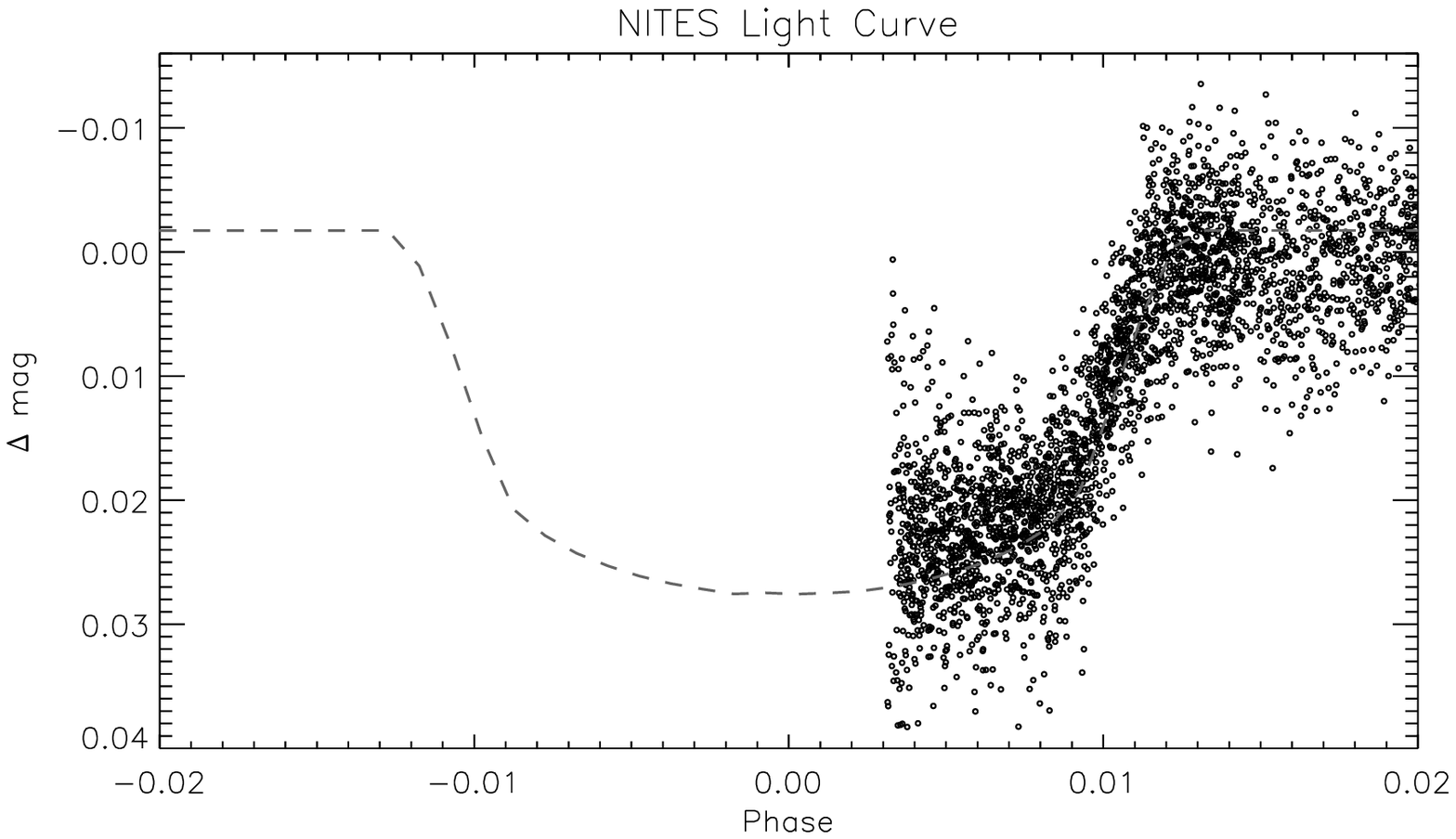}
\vspace{-1.8cm}
\caption{Follow-Up Photometry of Primary Eclipse. 
On the top, we show the time series photometric points acquired at the OED (\S \ref{oed}),
phased to the ephemeris from \S \ref{mcmc} and 
over-plotted with the best fit EB model described in \S \ref{eb} (red line).  We also show directly below
the residuals to the fit and the uncertainty of the individual photometric data points.
Similarly in the middle panel, we show the data from BYU and the residuals to the fit. 
In the bottom panel, we show the partial primary eclipse data acquired with NITES. 
Overplotted in grey is a model light curve in the R-band, 
shown only for guidance and comparison to the upper panels. 
The NITES data taken without a filter, and as such, we only utilise the NITES phtometry to derive the 
ephemeris, excluding it from the EB analysis (\S \ref{eb}).}
\label{figlcs}
\end{figure}

Because of the long period of the system, it is challenging to acquire a full eclipse on a single night
from a given location.  Thus, 
photometry from Observatori Esteve Duran (OED)
was performed on three nights, with a ST-9XE CCD camera attached to the
0.6m Cassegrain telescope and a 512$\times$512 pixels CCD. 
Each pixel is 20$\mu$m~$\times$~20$\mu$m, giving an image resolution of
$1.37\arcsec \times 1.37\arcsec$ on the CCD plane. An Optec I-band Bessel filter was used for the observations and the exposure time was
between 15 and 30 seconds. Photometry was extracted by performing
synthetic aperture photometry. The star GSC~2295-0229 was used as comparison.
 The flux of our target was divided by the flux of
the comparison star and differential
magnitudes was then obtained. The final light curve (shown in Fig.~\ref{figlcs}, top panel)
was obtained after binning of the
original photometric points with a typical magnitude scatter of 
$\sim$2--3 mmag.
The features in the OED photometry (at the end of the ingress and at mid-transit)
 are most likely due to the normalisation of the three individual partial eclipses that comprise
this light curve.  Based on the residuals between the data and the model (Fig.~\ref{figlcs}), these features in the light curve are not systematically affecting the fit to the light curve.

\subsection{Photometry--Optical: BYU}\label{byu}

The Brigham Young University (BYU)  0.91-m telescope is  a f/5.5 system located at the 
West Mountain Observatory in Utah, USA. 
It is fitted with a Finger Lakes PL-09000 CCD camera, 
a 3056$\times$3056, 12$\mu$m/pixel array that gives a field of view 
(FOV) of 25.2\arcmin$\times$25.2\arcmin\ \citep{Barth2011}.

 A near complete I-band
light curve of the primary eclipse was obtained on 2012 October
8. Pre-eclipse, ingress, flat bottom and egress was obtained. 
The resulting
I-band light curve consists of 2200  observations with a 17~second cadence, and  
 is presented in Fig.~\ref{figlcs}, middle panel.
The raw data were processed in the standard way.
After the instrumental signatures were removed, source detection and
aperture photometry were performed on all science frames using the
Cambridge Astronomical Survey Unit (CASU) catalogue extraction
software \citep{Irwin2001}. The optimal aperture radius was chosen
through empirical testing of several different sizes. The median
seeing of our observations was 3-4 pixels. We adopted a circular
aperture with a 14-pixel radius, several times the FWHM of our
relatively bright targets, to obtain the final instrumental magnitude
measurements. Two nearby stars of comparable brightness 
(GSC~2295-0229 and GSC N322212139) 
in the FOV of the detector were chosen as comparison objects for deriving
differential photometry. 
Other stars were considered as comparisons but were excluded
to minimise the scatter in the photometry. 

The total flux enclosed in the photometry
aperture for the reference objects was divided by the instrumental
flux of the target for each data point and then converted to
magnitudes. All the measurements were then normalised using the
out-of-eclipse portions of the light curve.

\subsection{Photometry--NIR: FLAMINGOS}\label{sececl}

\begin{figure}
\centering
\includegraphics[width=0.49\textwidth]{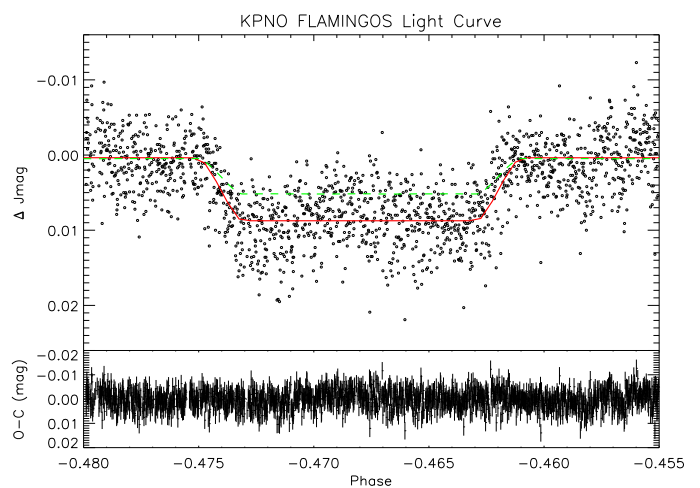}
\caption{Follow-up J-band Photometry of Secondary Eclipse.
We show the time-series photometry acquired at the Kitt Peak 2.1m telescope with
FLAMINGOS, superimposed is the best fit model (\S \ref{eb}; red line) 
from which we derive a temperature of 3922 $\pm$ 42~K for the M dwarf. 
Below, we show the residuals to the fit and the individual photometric uncertainties.
On the top panel, we also show a model light curve 
(green, dashed line) considering a cooler temperature for the M~dwarf (3350~K)
as expected from theoretical models (see discussion in \S \ref{disc}).  
This cooler temperature model light curve is significantly shallower 
than our best fit model (shown in red) and does not fit the depth of the secondary eclipse.  }
\label{fignir}
\end{figure}

We observed \eblm\ with the Florida Multi-Object Imaging
Near-Infrared Grism Observational Spectrometer (FLAMINGOS) in its
imaging mode mounted on the 2.1m telescope at Kitt Peak National
Observatory on the nights of 2012 October 28 and 29. Continuous
near-infrared (NIR) time-series photometry of the target
with the Barr
$J$-band filter was obtained on both
nights: the first night was dedicated to obtain out of eclipse
photometry; and during the second night the secondary eclipse of the
target was observed. The FLAMINGOS field-of-view is 20\arcmin
$\times$ 20\arcmin, which allowed the placement of the brightest
reference star in the same quadrant as the target. Our observing
strategy consisted in keeping the stars in the same position on the
detector, with no dithering. We defocused the telescope at the
beginning of the night, and then actively changed the focus to keep
from saturating the detector as the temperature of the telescope and
airmass of the target changed its FWHM. The 5 s science exposure
times ensured that no shutter correction was needed during the
reduction process of the images. Calibration frames were obtained on
both nights. Dome flats were obtained at the beginning of each night
with an exposure time of 8 s, as well as a series of night sky flat
field images (150 s). Dark frames of 5, 8, and 150 s were also
obtained.

We reduced the 1700 target frames in the standard manner: we first
subtracted the dark current frames from both the flat-field frames and
the target exposures. We then created a master flat-field frame and
normalised its average counts to $1$. Then, all the target frames were
divided by that master flat field frame, and in a further step bad
pixels and cosmic ray hits were deleted in all the target exposures.

We extracted the stellar counts of the target star plus two comparison
stars (GSC~2295-0229 and  GSC N322212139) 
by adopting aperture photometry. 
We chose the two comparison stars because of their brightness
and because no intrinsic flux variations in our photometric data were found in the two stars.
We used a self-written code for
properly centering the stars in the 30 pixel wide apertures. To determine
the average sky background value, we measured the count rates in a
ring around the star. We paid particular attention at masking out
faint stars present in those sky background rings. In the next step,
we subtracted the mean sky background values from the stellar count
rates and then calculated the final light curve by dividing the target
count rates by the sum of the fluxes of both comparison stars. In the
final step, we de-trended the light curve, thereby normalising the
out-of-eclipse flux to one, and calculated the Heliocentric Julian
date.
The middle of the secondary eclipse was affected by the non-linearity 
of the detector reached while the target was a its lowest airmass of 1.0.
When we exclude the affected section of the light curve
the measured depth of the secondary eclipse is consistent 
with the depth measured including all photometric points.

\begin{figure}
\centering
\includegraphics[width=0.49\textwidth]{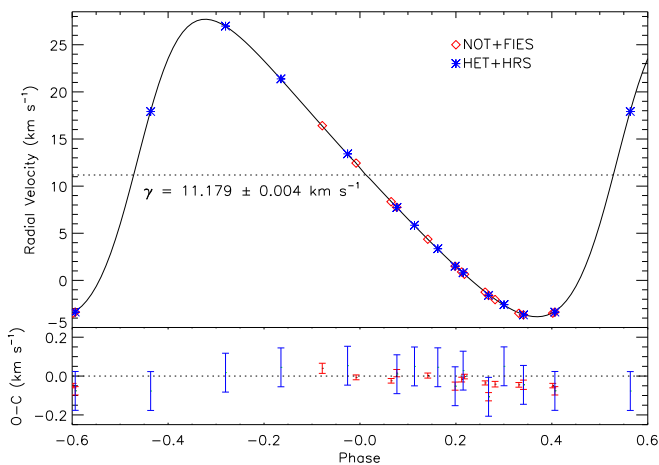}
\caption{Primary Radial Velocity Curve.  
We show the radial velocities describing the motion of the primary star around the
system's centre of mass.  The best fit model (\S \ref{m1}-\ref{eb}) is shown by the continuous (black) line with the systemic velocity given by the dotted line; 
the RV measurements acquired at the NOT are denoted by  red diamonds and  
the HET data are shown with  blue asterisks.  
The bottom panel shows the residuals to the fit and the uncertainties of the individual RVs.}
\label{figrvs}
\end{figure}

\subsection{Radial Velocities: NOT+FIES}\label{not}
We obtained follow-up spectroscopic observations to determine the
EBLM's orbital and stellar parameters.  \eblm\ was initially
observed using the FIbre-fed Echelle Spectrograph (FIES) mounted on
the 2.5-m Nordic Optical Telescope. In total, 15 usable spectra were
obtained between 2011 August 23 and 2012 August 14. FIES was used in
medium resolution mode (R = 46,000) with interlaced ThAr
calibrations, and observations were conducted using exposure times of
1200s (S/N $\sim$ 50) covering the wavelength range between 3630 and
7260 \AA. We used the bespoke data reduction package
FIEStool\footnote{http://www.not.iac.es/instruments/fies/fiestool/FIEStool.html}
to extract the spectra. An IDL cross-correlation routine was used to
obtain radial velocities (RVs) by fitting gaussians to the
cross-correlation functions (CCFs) of the spectral orders and taking
the mean. A template spectrum was constructed by shifting and
co-adding the spectra, against which the individual spectra were
cross-correlated to obtain the final velocities. The template was
cross-correlated with a high signal-to-noise spectrum of the Sun to
obtain the absolute velocity to which the relative RVs were
shifted. The RV uncertainty is given by RMS($\nu$)/$\sqrt(N)$, where
$\nu$ is the RV of the individual orders and N is the number of
orders.

Figure \ref{figrvs} shows a phase-folded orbital fit to the NOT (red diamonds) and 
HET observations 
(blue asterisks). 
The modelling to obtain the orbital parameters is explained in \S\ref{mcmc}, and 
the best-fit values are listed in Table~\ref{tableprop}.

\subsection{Radial Velocities: HET+HRS}\label{het}
Additional spectroscopic observation were obtained between October and November 2011 
using the High Resolution Spectrograph \citep[HRS;][]{Tull1998} mounted on the 9.2m 
Hobby-Eberly Telescope \citep{Ramsey1998}. The 316g5936 cross-disperser setting, 
together with a 2\arcsec optical fiber and a slit that provides resolving power of 
R $\sim$ 30,000 were used. Each observation was a 120s integration which yielded 
S/N $\sim$ 150 per resolution element.  Science observations were bracketed 
before and after with a ThAr hollow-cathode lamp exposure for wavelength calibration. 
The HRS data was extracted, reduced, and wavelength calibrated using a custom 
 optimal extraction pipeline written in IDL \citep{Bender2012}.  

We computed the CCFs for each epoch by the cross-correlation of fully 
reduced and calibrated spectra with a weighted G2 stellar template mask 
\citep{Pepe2002,Baranne1996} created using an NSO FTS solar atlas \citep{Kurucz1984}. 
 Radial velocities are determined by fitting the CCF with a gaussian \citep{Wright2013}. 
We find that the HRS RV measurements of $\sigma$ Dra, a well-known stable star, 
observed with the HET--HRS at the same setting over a period of 3 
months give an rms of $\sim$ 56 m s$^{-1}$. 
We therefore quote this as our formal measurement error. We note that this error, 
though higher than the error bars from photon noise, is more appropriate to use since the 
velocity precision is limited by instrumental effects, not by photon noise.

\section{Data Analysis}\label{analysis}

In this section, we describe the procedure by which we derive the physical properties of the
eclipsing system.  
Combined analysis of the observed radial velocity curve and photometric light curve of an eclipsing
binary provides the individual masses and radii of the component stars as well as their
temperature ratio and the absolute dimensions of the system.  
Given that \eblm\ is a single-lined eclipsing binary (see Fig.~\ref{figrvs}),
one additional parameter, in this case the mass of the primary star, 
is required to complete the solution \citep[e.g.,][]{Kallrath2009}.  

The analysis of a single-lined eclipsing binary is similar in many ways to a transiting planet system.
However, in the case of the single lined EB, the mass of the secondary component is not negligible. 
Therefore, the complete solution requires an interative approach to the analysis.
Derivation of the primary star radius from the eclipse light curve depends directly on the 
primary star mass through the semi-major axis.  However, the primary star mass is derived  
through a comparison to stellar evolutionary models using its radius, as well as its temperature and metallicity.
Below, we describe our iterative analysis procedure which has been applied
to \eblm\ until convergence of all the parameters was achieved.

\subsection{Stellar Characterisation: Primary star temperature and metallicity}\label{sme}

In order to derive the temperature, metallicity, and gravity of \eblm, we performed stellar
characterisation using the Spectroscopy Made Easy \citep[SME,][]{Valenti1996}
spectral synthesis code.  At the core of SME is a radiative transfer 
engine that generates synthetic spectra from a given set of stellar parameters.
Wrapped around the core engine is a Levenberg-Marquardt solver that finds the
set of parameters (and corresponding synthetic spectrum) that best
matches observed input data in specific regions of the spectrum.  The basic parameters 
used to define a synthetic spectrum are temperature (\teff),
gravity (\logg), metallicity ($[M/H]$), and iron abundance (\feh).  In order to
match an observed spectrum, we solve for these four parameters, plus the rotational broadening ($v~sin~i$)
of the star.  Our implementation of SME (described in Cargile, Hebb et al.\ in prep) varies
from the SME described in \citet{Valenti2005} in several ways.  
First, we used the ACCRE High-Performance Computing Center
at Vanderbilt University to run SME 300 times on the same input data but with a range of
initial conditions.   By allowing SME to find a best-fit synthetic spectrum from
a large distribution of initial guesses, we explore the full $\chi$-squared space 
and find the optimal solution at the global minimum.  In addition, we apply a line list
based on \citep{Stempels2007,Hebb2009} 
that includes more lines for synthesis, 
including the gravity sensitive Mg b triplet region.
Furthermore,  we use the MARCS model atmosphere 
grid in the radiative transfer engine, and we obtained the microturbulence ($v_t$) 
from the polynomial relation defined in \citet{Gomez2013b}.

We applied our SME pipeline to the two independent spectra obtained with 
the different telescope and instrument
configurations (\S \ref{not} and \ref{het}).  We shifted and stacked four NOT observations
obtained in the 2011 observing season to generate a single moderate signal-to-noise spectrum 
for stellar characterisation.  We also shifted and stacked the 15 HET observations into a single,
high signal-to-noise spectrum to use for an independent stellar characterisation analysis. 
We applied our pipeline to each combined spectra while allowing all 5 parameters to vary freely. 
The results of the SME analysis are  presented in Table~\ref{tablespec}.
The two datasets have different resolutions and signal-to-noise properties, but we derive
the same stellar parameters from both spectra.   
However, we were only able to derive a robust \vsini\ from the NOT spectrum given its higher resolution and signal-to-noise ratio, which we found to be 5.87 \kms.  
For the final stellar parameters of \eblm, we adopt a weighted mean of the results from the two independent analyses.

The uncertainties on each set of parameters include both statistical and systematic uncertainties added in quadrature.  
The formal 1-$\sigma$ errors are based on the $\Delta \chi^2$ 
statistic 
for 5 free parameters derived from our distribution of 300 final solutions. 
To derive the systematic uncertainties, we compared the results 
of four independent stellar characterisation analyses  and 
report the mean absolute deviation of our results for the case of the planet-hosting star
WASP-13 \citep{Gomez2013b}. 
The systematic errors we derived  
are $\sigma = \pm 48$~K in \teff, $\sigma = \pm 0.07$ in \logg, and $\sigma = \pm 0.03$ in \feh.

\begin{table}
\caption[]{Spectroscopically-determined Properties of Primary Star}
\begin{center}
\begin{tabular}{lccc}
\toprule \hline \\
   & HET & NOT  & Weighted Mean\\
\hline\\
\tprim & 5962 $\pm$ 82 & 5961 $\pm$ 72& 5961 $\pm$ 54 \\ 
\feh	& -0.41 $\pm$ 0.06 &  -0.39 $\pm$ 0.05 & -0.40 $\pm$ 0.04 \\
\logg& 4.02 $\pm$ 0.10  & 4.15 $\pm$ 0.10 & 4.09 $\pm$ 0.07 \\
\bottomrule
\end{tabular}
\end{center}
\label{tablespec}
{\footnotesize The individual errors on the HET and NOT results include a systematic uncertainty as defined in \citet{Gomez2013b}.}
\end{table}

\subsection{Ephemeris, orbital properties, and K$_1$}\label{mcmc}
 
We applied a Markov-Chain Monte Carlo (MCMC) analysis simultaneously to all our available data: the   
SuperWASP photometry, the higher precision, optical photometry from NITES, OED and BYU, and 
the KPNO NIR secondary eclipse light curve, together  
with the NOT and HET radial velocity measurements.  A detailed description
of the method is given in \citet{Cameron2007} and \citet{Pollacco2008}, as is typically applied 
for transiting planetary systems. 
The star--planet system represents an example of a single-lined eclipsing binary with 
an extreme mass ratio ($q \ll 1$), and
our model follows that described by \citet{Mandel2002} which assumes that the mass of the
secondary component (e.g., the planet or lower mass star) is significantly lower than 
the mass of the primary star ($M_{2} \ll M_{1}$).  
Because this assumption on the mass ratio is no longer valid for in the case of \eblm, 
which has an secondary mass $\sim$0.2~\msun, we do not utilise this MCMC analysis to measure
the absolute dimensions and masses of the binary stellar components.  Instead
we use this well-tested code to measure the properties
that can be directly measured from the complete dataset with robust uncertainties, namely:
the orbital period $\porb$, the time of mid-primary eclipse $T_0$, the eccentricity $e$,
the argument of periastron $\omega$, the radial velocity semi-amplitude K$_1$,  
the centre-of-mass velocity of the system $\gamma$, and depth of the secondary 
eclipse ($\Delta$ F$_{\rm sec}$).  In the case of the orbital 
geometry, $e$ and $\omega$ are derived from the 
Lagrangian elements \secos\ and \sesin\ that are tightly constrained  
by the well-sampled RV curve, and the secondary eclipse light curve.  
It must be noted that the derived value of $e$= 0.3098 is very close to the eccentricity (0.308)
 at which the rotational angular motion of the star at periastron 
is highest as compared to the mean orbital motion, as defined by \citet[][after Eq.~48]{Hut1981}.
Furthermore, $\gamma$ was allowed to vary independently for the different RV 
datasets to allow for systematic offsets between the instruments/telescopes/observing runs.  
All the values derived from the MCMC analysis are given in Table~\ref{tablepreeb}.

\subsection{Primary star mass determination} \label{m1} 

\begin{figure}[h]
\centering
\includegraphics[width=0.49\textwidth]{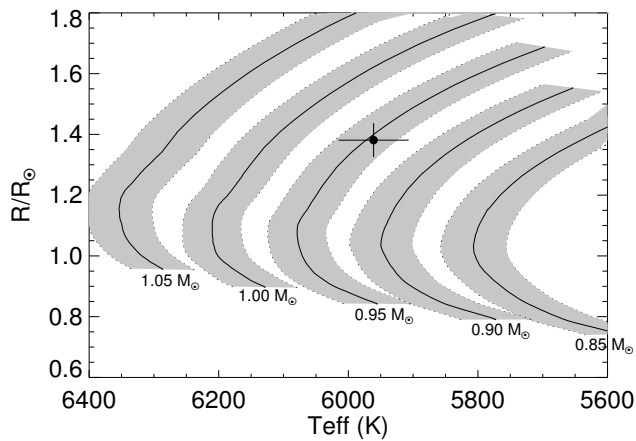}
\caption{Mass tracks from Y$^2$ models of \citet{Demarque2004} interpolated
to \feh=$-0.4$ (solid lines) for 5 different masses.  
The gray bands around each track indicate
the range of properties at those masses which are consistent with the uncertainty
on the metallicity.  The values for the radius and temperature of the primary star 
are plotted as the solid circle with uncertainties.  Through this comparison
with the stellar evolution models, we find the mass of the primary star to be
$0.945\pm0.045$ \msun.
}
\label{figm1}
\end{figure}

The spectroscopically-determined stellar parameters derived above (\S \ref{sme})
and the radius of the primary star are 
 compared with subsolar metallicity 
Yonsei-Yale (Y$^2$) stellar evolutionary models  \citep[see Fig.~\ref{figm1};][]{Demarque2004}
to constrain the mass of the primary star.
We generate a highly dense grid of mass tracks by interpolating the
Y$^2$ models in temperature, radius, and metallicity.  
The spectroscopic properties of the primary star (e.g., \teff\ and \feh)
are independent from the primary mass, and as such, are kept fixed. 
Since our target is an eclipsing binary, we are able to measure the radius of
the star from the combined analysis of the radial velocity curve and light curve 
(\S \ref{eb}).  Therefore, we compare the radius of the star instead
of the luminosity or spectroscopic gravity, as is typically done
for single stars, because the stellar radius is a more precise 
and direct observational quantity for our system.

We then find the range of stellar masses that are consistent with the derived stellar
parameters including their uncertainties.  
Figure~\ref{figm1} shows a modified Hertzprung-Russel diagram comparing the derived properties
of the primary star to theoretical mass tracks interpolated to \feh=$-0.4$ dex.
The figure shows a series of mass tracks at exactly the most
probable metallicity of the star while the gray bands around each track indicate
the range of properties at those masses which are consistent with the uncertainty
on the metallicity.

It is important to note that while the spectroscopically derived \teff\ and \feh\ are 
independent of the mass and are kept constant, 
the radius derived from the eclipsing binary model is not.
Thus, we derive the primary star mass from the stellar models and then perform the
eclipsing binary model several times in an iterative fashion until 
all the quantities are consistent and 
agree within their uncertainties.  
We iterate until the primary mass derived from the theoretical models using a particular primary radius
 is consistent with the primary mass that is input into the EB model
from which that same radius is derived.
We first estimate the radius of the star based
on a light curve solution using the transiting planet approximations (see \S \ref{mcmc}). 
Although the planet approximations are
 not valid in the case of EBLMs, we use this 
preliminary radius (1.273 $\pm$ 0.028 \rsun) only as a starting point for our analysis  
deriving an initial primary star mass.
 This resulting primary mass  is then fed back
into a full eclipsing binary model (\S \ref{eb}) to derive a new primary star radius.    
After two iterations of this procedure, we achieved convergence.   In addition,
our independently derived spectroscopic gravity was found to be consistent with
the gravity derived from the mass and radius of the primary star after convergence.
The final primary star mass is given in Table~\ref{ebproperties}.  The uncertainties 
in the primary mass are obtained
from the range of Y$^2$ mass tracks that are consistent with the derived properties.
We derive an approximate age for the system from the models of $\sim 9.5$~Gyr.  This is reasonable given
the low gravity and low metallicity.

However, as a sanity check on the relatively old age,
we calculate the rotation period of the primary star and compare it to the 
expected rotation period if the system has had enough time to sychronise.
Since the binary is eccentric, this is called the pseudo-psynchronous period.
We find the period of the primary star to be, $P_{\rm rot} = 11.88$~d, using 
the spectroscopically-determined \vsini\ and the primary radius and assuming the spin
axis of the primary star is aligned with the orbital axis of the secondary star.  While not always
a reasonable assumption for transiting planet systems, this is reasonable for binary stars. 
We then calculated an estimate of 
the expected pseudo-synchronous period as defined by \citet[][Eq.~45]{Hut1981}
for eccentric stellar binaries assuming a constant configuration for the binary and its stellar components 
to be $P_{\rm pseudo} = 11.42$~d.
Considering the uncertainties in the measurements, the 
rotation period of the primary is consistent with being synchronized at periastron.  
In addition, a close binary such as \eblm\ is expected to have synchronized its rotation in $\sim$4 Gyr,
following \citet{Zahn1977}.  Thus these additional constraints indicate that the age of \eblm\ 
is older than $\sim$4 Gyr, in agreement with the age derived from the evolutionary models.

%&From MCMC analysis &&\\
\begin{table}[h]
	\caption[]{Ephemeris and Orbital properties of \eblm}
	\begin{center}
	\begin{tabular}{lcc}
		\toprule \hline \\
		   & Value   & Units \\ 
		\hline\\
		 & From MCMC: & \\
		\hline\\
		T$_0$& 6023.26988 $\pm$ 0.00036        & days$^{\dagger}$ \\
		\porb& 14.2769001 $\pm$ 0.0000067 &   days \\
		$\Delta$ F$_{\rm sec}$&  0.00737 $\pm$ 0.00024 & \\
		$e$&   0.3098 $\pm$ 0.0005     &       \\
		$\omega$&      278.85 $\pm$ 1.29    &       degrees \\
		$\gamma$       & 11.179 $\pm$ 0.004    & \kms \\
		K$_1$&  15.84 $\pm$ 0.01 & \kms \\
		\hline\\
		& From EB formulae:  & \\ \hline \\
		K$_2$&	80.3 $\pm$ 1.5 &\kms\\
		q&	0.1968 $\pm$ 0.0035& \\
		$a \sin{i}$& 25.808 $\pm$ 0.387 & \rsun \\
\bottomrule
\end{tabular}
\end{center}
\label{tablepreeb}
     {\footnotesize $^\dagger$ Heliocentric Julian Date -- 2\,450\,000 }  
\end{table}

\subsection{Mass Ratio and Semi-major Axis}\label{qasini}

Using the orbital elements and $K_1$ derived in \S\ref{mcmc}, and $M_1$ derived in \S\ref{m1}, 
we solve for $K_2$ from:  
\ensuremath{M_1 \sin^3{i} = 1.036149 \times 10^{-7} (1-e^2)^{3/2}(K_1+K_2)^2K_2\porb}, 
 which includes updated 
values for the Solar radius and for the heliocentric gravitational
constant \citep[][and references therein]{Torres2010}.
The units of the primary mass is solar mass, the $K_1$ is in \kms, 
and the orbital period \porb\  
is in days. 
The 1$-\sigma$ uncertainties are given by the extremes allowed by the errors in the 
parameters from which $K_2$ is derived. 
Once $K_2$ has been computed, the mass ratio is directly obtained from $q = K_1/K_2$; 
the semi-major axis of the orbit (in solar radii) as a function of the 
inclination is derived from 
\ensuremath{a\sin{i} = 1.976682 \times 10^{-2} (1-e^2)^{1/2}(K_1 + K_2) \porb}. 
Table~\ref{tablepreeb} contains these derived values and their 1$-\sigma$ uncertainties.

\subsection{Eclipsing Binary Modelling: Inclination, Stellar Radii and Secondary Temperature}\label{eb}

\begin{table}
	\caption[]{Parameters measured from the Light Curve Modelling}
	\begin{center}
		\begin{tabular}{lccc}
			\toprule \hline \\
			   & WD2010 & PHOEBE & Weighted Mean\\
			\hline\\
			$i (^\circ)$&		89.083$\pm$0.037 	& 89.13$\pm$0.27 	& 89.084  $\pm$ 0.037 \\
			$r_1$&			0.0532$\pm$0.003	& 0.0535$\pm$0.0028 &  0.0534 $\pm$ 0.0021\\
			$r_2$&			0.0081$\pm$0.0006	& 0.0081$\pm$0.0005	&  0.0081 $\pm$ 0.0004\\
			T$_1$/T$_2$&		1.531$\pm$0.012	&  1.509$\pm$0.012 & 1.520 $\pm$ 0.009 \\ 
			\bottomrule
		\end{tabular}
	\end{center}
	\label{tablelc}
\end{table}

\begin{table}
	\caption[]{Physical properties of \eblm} \label{ebproperties}
		        \begin{center}
			\begin{tabular}{lccc}
			\toprule
		\hline\\
		   & Value   & Units \\ 
		\hline\\
		$a$ &25.811 $\pm$ 0.387 & \rsun \\	
		M$_1$&	0.945 $\pm$ 0.045& \msun\\  
		M$_2$&  0.186 $\pm$  0.010	& \msun \\
		\rprim & 1.378	$\pm$ 0.058& \rsun \\
		\rsec & 0.209	$\pm$ 0.011& \rsun \\
		\logg &4.14 $\pm$ 0.04 &  dex\\
		\loggs &5.07 $\pm$ 0.05 &  dex\\
		\tsec\ & 3922$\pm$ 42 & K \\
		$L_1$ &  2.154  $\pm$  0.197  & L$_\odot$ \\
		$L_2$ & 0.009  $\pm$  0.001 & L$_\odot$ \\
 \bottomrule
                 \end{tabular}
		         \end{center}
			         \label{tableprop}
			 \end{table}

Because the MCMC model assumes (a) a secondary component with a negligible mass  
that has (b) an opaque dark surface, and requires that (c) the secondary
radius is less than 10\% of the stellar radius \citep{Mandel2002},   
it is necessary to use standard eclipsing binary modelling tools to derive
primarily the inclination of the orbit, 
the stellar radii, and the temperature of the secondary component.
We applied two techniques both based on the widely used Wilson-Devinney code
 (hereafter WD; \citealt{Wilson1971}). 
The WD code allows to fit simultaneously the light and the radial 
velocity curves of eclipsing binaries deriving consistent parameters for all the data.  
For this analysis we used the OED, BYU, and KPNO light curves, and 
the radial velocities of the primary.  
The WASP and NITES photometry 
were excluded from these analyses because WD calculates each light curve
based on model atmospheres and observed pass-bands;  
this photometry was obtained with non-standard filters. 
We adopted the parameters derived in the previous sections and kept them fixed in the EB modelling, namely:
$q$, $e$, $\omega$, \porb, $T_0$, \tprim, and \feh.  
It must be noted that the EB model depends directly on the primary mass (\S \ref{m1}),
as $q$ is a needed input.  Thus, the derived primary radius resulting from the
EB modelling is fedback iteratively into the primary mass determination until the derived 
primary mass is consistent with the primary mass used in the EB model. 
Spin--orbit synchronisation was assumed for the two components. 
Furthermore, rotational effects for such a long period binary are not expected to be significant.
The reflection albedos were fixed at the value 0.5, appropriate for stars with convective envelopes \citep{Rucinski1969}, 
and the bolometric gravity darkening exponents were set to
0.4 for the primary and 0.2 for the secondary, following \citet{Claret2000b}.

As shown in Table~\ref{tablelc}, from each independent EB analysis, we derived the inclination angle of the orbit ($i$), the fractional radii ($r_j= R_j/a$), and the temperature ratio (\tprim/\tsec).  We then combined
these two sets of results via a weighted mean to obtain their adopted values.   These parameters
depend solely on the light curves, and as such are independent from the mass ratio and \asini.   
As mentioned above,
because the primary star mass depends on the stellar radius, we iterated until convergence:
updating the stellar radius 
in the mass determination, then deriving updated mass ratio and orbital separation, and consequently,
new stellar radii.  

The final EB solution indicates that 
the two stars are spherical as expected for long period binaries. The primary eclipse is annular, while the secondary is total, and the totality phase lasting about 3.1 hours. This provides the opportunity to constrain the effective temperature and metallicity 
of the primary component from high signal-to-noise ratio spectra not contaminated by the secondary component.
In conjunction with the results from \S\ref{sme}-\ref{qasini}, we derived the final physical properties of the orbit and stellar components, as given in Table~\ref{tableprop}.

\subsubsection{Wilson-Devinney (2010) Modelling}\label{wd}
The first EB modelling of the light and radial velocity curves was performed using the 2010 version of the WD code. 
The main parameters adjusted were the orbital inclination ($i$), the pseudo-potentials (\potp\ and \pots), 
the temperature ratio (\tprim/\tsec),
the semi-major axis ($a$), 
the systemic velocity ($\gamma$), the primary luminosity ($L_1$) for each bandpass (i.e., the light ratio), and the time of periastron passage (to account for variations of the times of eclipses).
Initially, we tried to also fit for $e$ and $\omega$; however the solution diverged given that the parameters
are highly correlated.  Based on the method of multiple subsets described by \citet{Wilson1976},
we fit for $e$ and $\omega$ independently of the other parameters, and obtained values for $e$ and $\omega$
that were consistent with those derived in \S\ref{mcmc}.
Emergent intensities used in the program were taken from model atmospheres described by \citet{VanHamme2003} and limb darkening coefficients where computed from \citet{VanHamme1993} as implemented in the WD code.
The coefficients were dynamically adjusted according to the current effective temperatures and surface gravities of the stars at each iteration. 

In the case of this first method, the convergence in the final fit was considered to have been achieved when the corrections to the elements were smaller than the internal errors in three consecutive iterations. This procedure was repeated five times and the solution with the smallest $\chi^2$ value was chosen as our best solution. Observational weights for each light curve were adjusted according to their residuals using a preliminary fit and scaled to get a $\chi^2 \sim 1$ for the light curves. The observational weights for the radial velocities were also scaled accordingly. In order to provide more realistic uncertainties for the geometric and radiative parameters than the internal errors estimated by WD, we continued the iterations in our adopted fit beyond convergence for another 200 steps, and we examined the scatter of those 200 solutions. We adopted the larger of the estimates for the parameters.
Third light was also tested for but was always found to either converge toward negative (unphysical) values, or to be roughly consistent with zero. 
The depth of the secondary eclipse measured from this WD analysis is 0.00741 $\pm$ 0.00001, consistent
with both the MCMC and {\sc PHOEBE} measurements.

\subsubsection{PHOEBE modelling}\label{phoebe}

The second EB analysis was performed using the WD-based code {\sc PHOEBE} \citep{Prsa2005}.  
Firstly, we made use of the scripter capability of {\sc PHOEBE} and cluster computing to sample a large parameter space.
The parameters sampled were the inclination of the orbit ($i$), and the stellar radii via the potentials 
(\potp, \pots), as defined for eccentric orbits by \citet{Wilson1979}. 

Utilising the Vanderbilt University Advanced Computing Center for Research and Education (ACCRE), 
we randomly selected values between $15.5 \leq \potp < 23.5$, $22.0 \leq \pots < 32.0$, 
and $88.2 \leq i < 90.0$.  
For each combination of $i$, \potp, and \pots, we calculated $a$ from \asini\ (\S\ref{qasini}) and the chosen $i$.  
To make this computationally expensive process more efficient while mapping the space around the minimum 
sufficiently well, after the first 10\,000 points,
we narrowed the sampled parameter range around the solution with the lowest $\chi^2$. 
The total number of different parameter combinations sampled was 36972.  
The calculation of $\chi^2$ includes the radial velocity curve and the follow-up light curves.  
It must be noted that the sampled parameters ($i$, \potp, and \pots) 
depend solely on the light curves.  However, the stellar radii depend on the absolute dimension of the 
binary orbit determined by $a$ (semi-major axis) which is derived from \asini\ (see \S\ref{qasini}), 
that depends on the RV curve, and the light curves via $i$. 
This allowed us to create a multi-dimensional $\chi^2$ map, and ensuring that our best solution 
(i.e., with the lowest $\chi^2$ value)  
corresponds to the global minimum. 
Furthermore, we are able to define  the 1-$\sigma$ uncertainties around the best solution 
from the $\chi^2$ map and to assess the correlations between the 
parameters \citep[e.g.,][]{Gomez2012}.  

In order to derive the temperature ratio (\tprim/\tsec), we included the secondary eclipse light curve
(\S\ref{sececl}) in the dataset modelled with {\sc PHOEBE}. We fitted primary and secondary eclipse light curves simultaneously to derive the best solution.  We obtained fractional radii and an inclination that are in agreement within 1$-\sigma$ with the result from the best ACCRE solution, and a temperature ratio of 1.509 $\pm$ 0.012, which in combination with the \tprim\ derived in \S\ref{sme} renders \tsec\ = 3950~K.
The depth of the secondary eclipse derived from our {\sc PHOEBE} analysis is 0.0074 in flux, agreement with the measurements from the WD and MCMC analyses.

\section{Discussion}\label{disc}

\begin{figure*}
\centering
\includegraphics[width=0.495\textwidth]{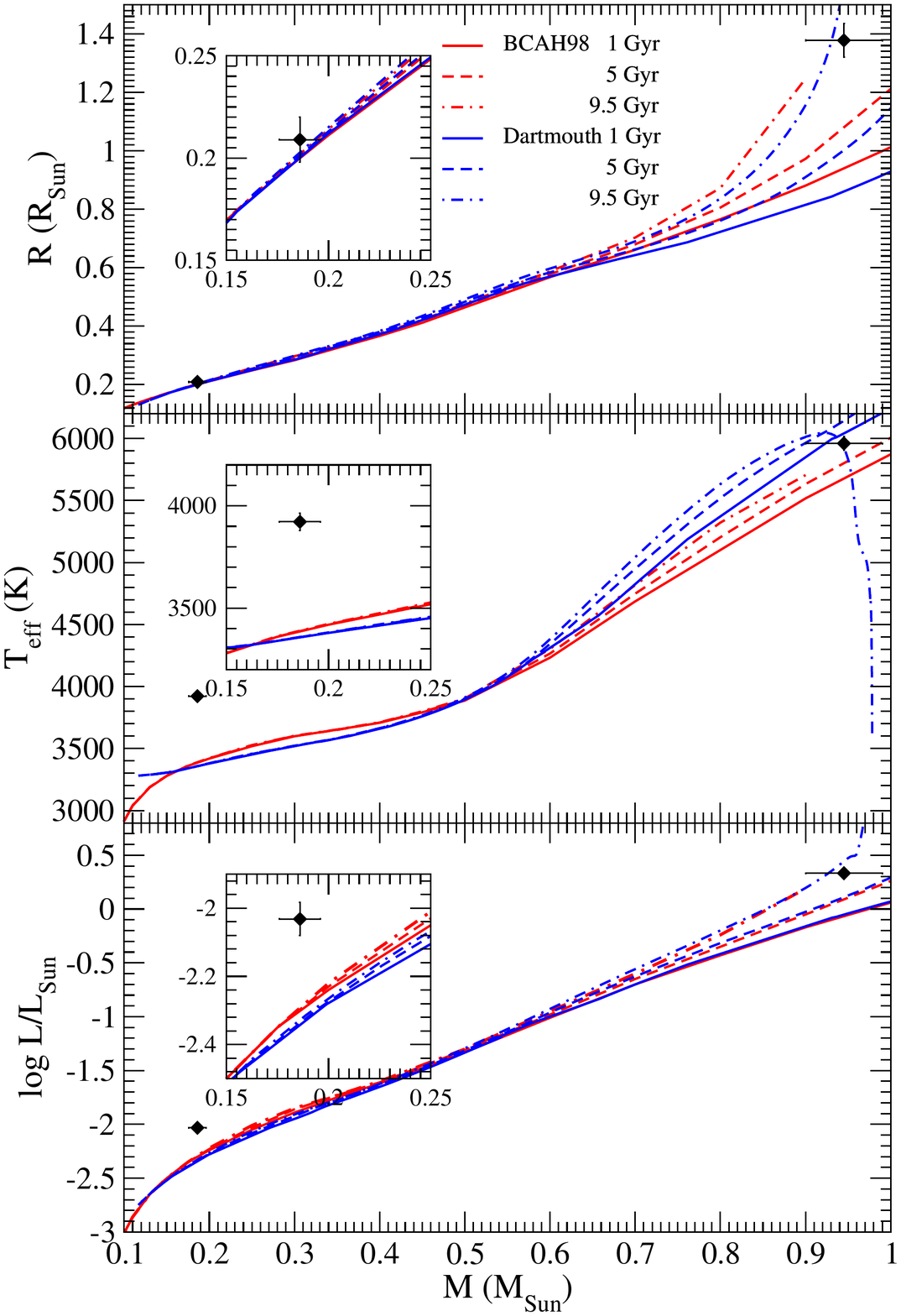}
\includegraphics[width=0.42\textwidth]{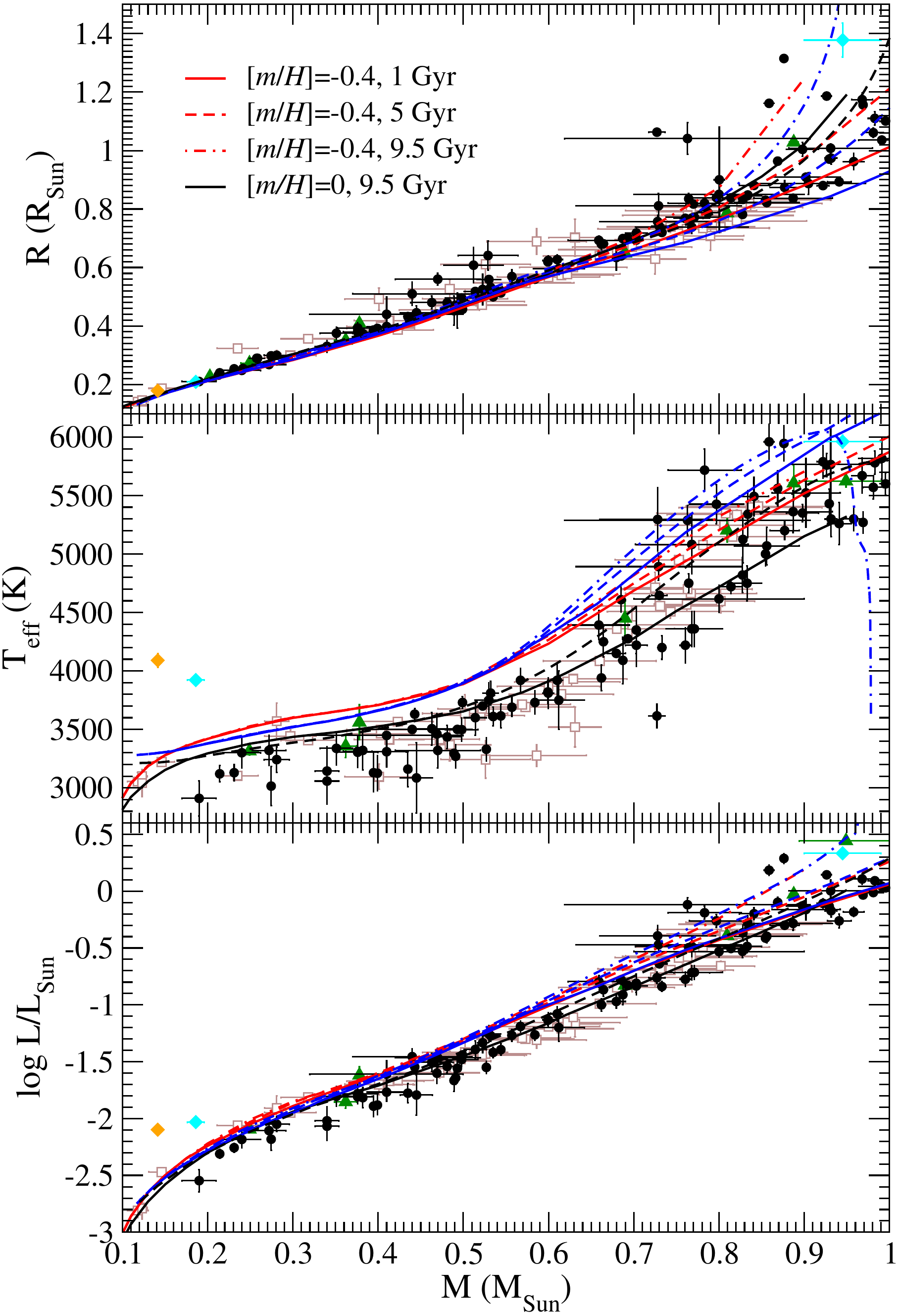}
\caption{Comparison to stellar evolution models and other direct measurements.
On the left, our measurements of stellar radius (top), effective temperature (middle), and
luminosity (bottom) for both components of \eblm\ are shown as black diamonds against the 
stellar evolution models of \citet[][in red]{Baraffe1998} and \citet[][in blue]{Dotter2008},
interpolated at a metallicity of $-0.4$ dex for different ages (solid line for 1 Gyr; dashed line 
for 5 Gyr; dash-dotted line for 9.5 Gyr).  The insets highlight for clarity the area around 
the measurements for the M dwarf.
On the right, we show the same measurements of \eblm\ (cyan-filled diamonds) and compare them to 
other measurements of M dwarfs: from double-lined eclipsing binaries (black-filled circles),
from {\it Kepler} eclipsing binaries with circumbinary planets (green-filled triangles), and with interferometric
radii (brown, open squares). 
Apart from \eblm, the only other M~dwarf in an single-lined EB with a 
measured temperature (KIC 1571511) is marked by the yellow diamond.
See the text for the references of the measurements included in this figure. 
The model evolutionary tracks are the same as on the left panel, 
except for the black lines which show for comparison the models of \citet[][solid]{Baraffe1998} 
and \citet[][dashed]{Dotter2008} for solar metallicity at 9.5 Gyr.  
}
\label{models}
\end{figure*}

Table \ref{tableprop} presents our final physical properties for the system and its stellar 
components which in Figure~\ref{models} are compared to two sets of stellar evolution models:
BCAH98 \citep{Baraffe1998} and DSEP \citep{Dotter2008}.   
Both sets of evolutionary models have been
interpolated to the metallicity of the EBLM ($-0.4$ dex). 
It must be noted that the BCAH98 models for the estimated age of our system (9.5 Gyr) 
only go up to masses of 0.9 \msun, because stars of higher mass at this metallicity beyond this age 
have evolved off the main sequence.
The differences between both set of models could be due to the different mixing-length parameters used in the readily available theoretical calculations  ($\alpha_{\rm BCAH98}$ = 1.0; $\alpha_{\rm  DSEP}$ = 1.938).  
In general, a more efficient convection 
(i.e., higher mixing-length $\alpha$) causes the overall radius to decrease
and the temperature to increase to maintain the same luminosity. 

As shown on the left side of Fig.~\ref{models}, the 9.5-Gyr DSEP model
reproduces very well our measurements for the primary radius, temperature and thus luminosity
for the derived mass of the primary.  
Given that the primary mass and the system's age are derived directly from the Y$^2$ models (\S \ref{m1}),
the intersection of the primary radius, temperature and luminosity with the 9.5-Gyr DSEP isochrone denotes
good agreement between these models in this mass regime.  
More specifically, our measurements of the primary radius and luminosity fall within their
uncertainties on 
the 9.5 Gyr DSEP isochrones in the mass-radius and mass-luminosity planes, respectively, 
and are not consistent with the ones at 1 and 5 Gyr.
In the temperature-mass plane, our measurement of \tprim\ is consistent within uncertainties with all three 
DSEP isochrones, and with the 5-Gyr BCAH98 isochrone.  

While both sets of models are able to reproduce the radius of the M dwarf  given its measured mass, its measured \tsec\ is significantly hotter than that predicted by the theoretical 
isochrones.  
Consequently, the observed luminosity is also well above the predicted one. 
Due to the slow evolution of low-mass stars, 
our measurement of the secondary radius is consistent with both sets of models from 1 to 9.5 Gyrs.  
At these low masses, the BCAH98 and the DSEP models 
are indistinguishable in the mass-radius plane. 
Both sets of models predict a star with the mass of the secondary to have
a temperature of $\sim$3350~K, which is $\sim$600~K cooler than 
our measured \tsec.  
Figure~\ref{fignir} compares our best fit light curve model (in red) for a 3922~K M~dwarf and 
that of a 3350~K (in green) as predicted by stellar evolution models.  
Our data is not consistent with the cooler temperature model light curve.  
Below we discuss different phenomena/effects that could impact our measurement of the secondary
effective temperature, by
either affecting our measurement
or by providing a source of heating additional to fusion in the M-dwarf interior. 
Any additional heating mechanism
must contribute $\sim 10^{31}$~ergs/s, which is about 70\% of
the energy produced by fusion for the 3350~K version of this star.  

\begin{figure}[!ht]
\centering
\includegraphics[width=0.495\textwidth]{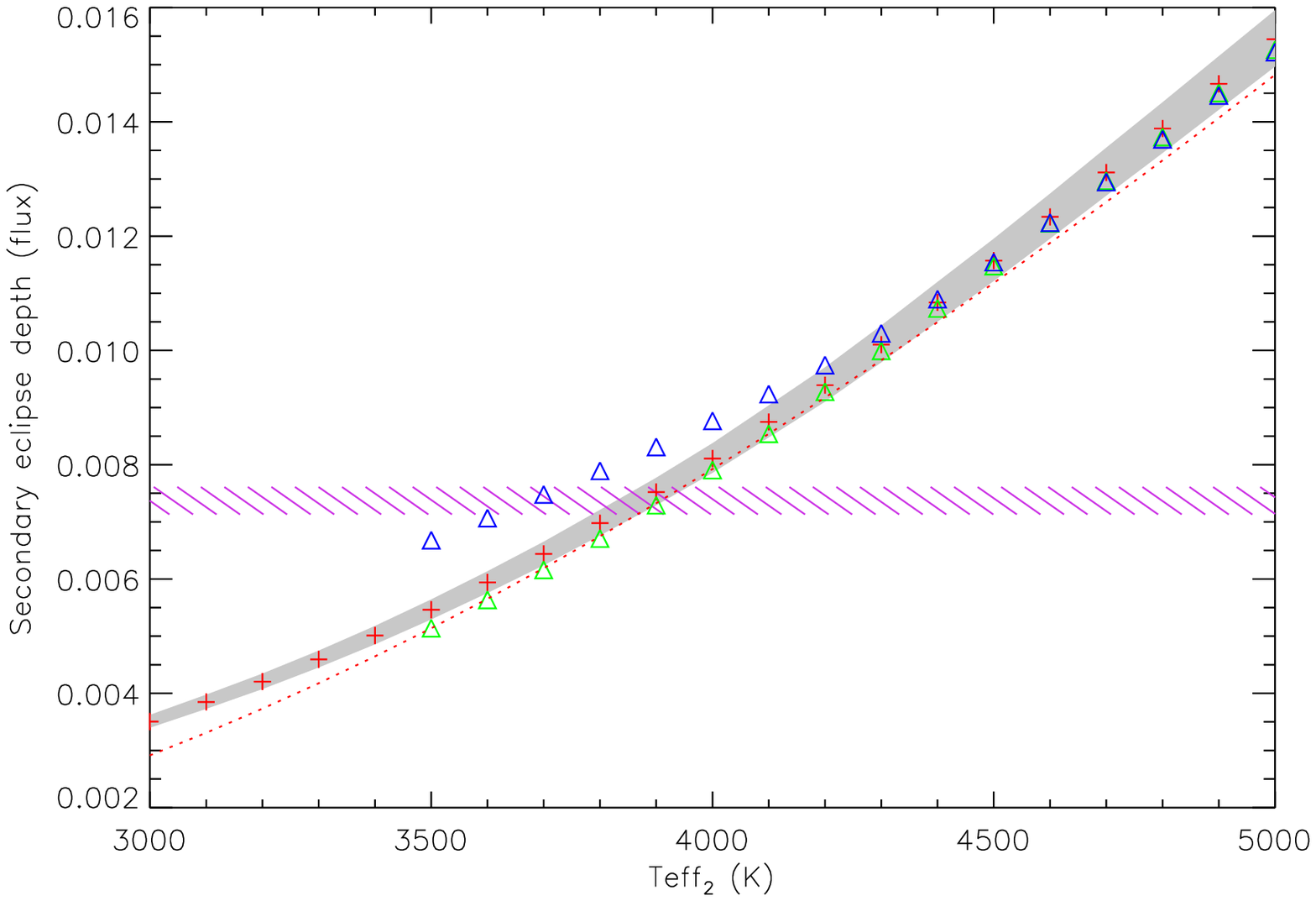}
\caption{Expected secondary eclipse depth based on the latest PHOENIX stellar atmospheres
from \citet{Husser2013}.
The magenta-lined area marks the measured depth of the secondary eclipse from our NIR light 
curve, 0.00737 $\pm$ 0.00024 in flux. 
The red crosses denote the expected secondary eclipse depth when assuming \feh\ = $-0.5$ dex,
\tprim\ = 6000~K, \logg\ = 4.00 dex, \loggs\ = 5.00 dex, and no alpha element enhancement,
and the grey area is considering an uncertainty in the primary temperature of 100~K. 
The triangles are similar to the ceslirosses, except that they are for different alpha enhancement levels
for both components: 
the blue triangles are for $\alpha$ = $+1.2$ dex, and the green triangles are for $\alpha$ = $-0.2$ dex. 
The red dotted line is the expected eclipse when using black bodies instead of the model atmospheres. 
Even the highest alpha element enhancement considered by the PHOENIX model
atmospheres is not enough to 
reconcile the secondary eclipse depth for \tsec\ expected from evolutionary models with 
our measurement. 
}
\label{atms}
\end{figure}

\begin{itemize}
\item {\bf Treatment of model atmospheres in EB modelling}: 
Using the most up-to-date model atmospheres for cool stars
\citep[PHOENIX;][]{Husser2013}, we investigate and verify that our measurement of the \tsec\ 
from the EB modelling is consistent with the latest stellar atmospheres, 
since the underlying model stellar atmospheres used by both PHOEBE and WD are those of Kurucz
\citep[][and references therein]{VanHamme2003}.
These PHOENIX atmospheres extend to very low effective temperatures (2300~K),
and include the wavelength range of our secondary eclipse ($\sim$1.1-1.4$\mu$m). 
We were not able to use the synthetic spectra of the AMBRE project, based on the MARCS model
atmosphers, because
their cut-off wavelength is 1.2$\mu$m \citep{deLaverny2012}.
Because of the discrete grid points calculated by \citet{Husser2013}, for 
this test we chose the closest values 
to our derived physical properties.  Thus, we utilised: \feh = -0.5 dex for both components 
(with no alpha elements enhancement; $\alpha$ = 0.0 dex),
for the primary \tprim\ = 6000 $\pm$ 100~K and \logg\ = 4.0 dex, and for the secondary, \loggs\ = 5.0 dex.
In order to compare our measurement of the eclipse depth to that expected by model atmospheres,
we integrated the stellar atmospheres over the transmission function of the Barr J-band filter of 
our NIR observations, and scaled the model atmospheres by the measured stellar radii. 
As shown in Fig.~\ref{atms}, 
we tested temperatures for \tsec\ between 2300 and 5000~K, in 100~K intervals, and 
confirm that the measured secondary eclipse depth (0.00737 $\pm$ 0.00024 in flux) is
consistent with the PHOENIX model atmospheres for a secondary star of $\sim$3900~K.

\item{\bf Determination of primary temperature and metallicity:} 
Given that it is \tprim/\tsec\ that is measured from the relative depth of the
eclipses and the spectroscopic \tprim\  and
that we adopt the metallicity of the primary as that of the M dwarf, 
we discuss the possibility that the stellar characterisation of the primary star is not accurate. 
If \tprim\ were cooler, consequently so would \tsec. 
However with the measured temperature
ratio, the primary temperature would need to be $\sim$900K lower
in order to reconcile \tsec\ with evolutionary models.  
Similarly, the metallicity of the primary star (and thus of the M dwarf) would need to be 
between $-1.5$ and $-2.0$ dex for our measured \tsec\ to be consistent.  
Both of the effects  (i.e., cooler \tprim\ and/or a more metal poor system) would imply the 
primary star to be lower mass than what we have estimated in \S\ref{m1}, 
and thus in turn the derived \tsec\ would still likely be higher than expected.  
The observed spectra are not consistent which such cool temperature nor such a low metallicity for the primary. 
Furthermore, our spectroscopic analysis of the two available spectra were done independently
and render results within 1-$\sigma$ of each other (see Table~\ref{tablespec} and \S \ref{sme}).
Although some studies have found that SME measurements 
show strong correlations between effective temperature, metallicity and \logg\ \citep[e.g.,][]{Torres2012}, 
our primary star is cooler than 6000~K, above which these systematics start to appear.
The fact that our spectroscopically-determined \logg\ is consistent with the \logg\ derived
from the EB modelling indicates that the spectroscopically-derived temperature and metallicity 
are not subject to the systematics problems outlined in \citet{Torres2012}.  
Thus, we conclude that the \tsec\ discrepancy is not likely due to an inaccurate stellar characterisation
of the primary star, and that the system is not sufficiently metal-poor 
to reconcile \tsec\ with the evolutionary models.

\item {\bf Alpha element enhancement}: Our stellar
characterisation of the primary star and the model atmospheres used in the EB modelling assume
solar abundances of the alpha elements. 
 Thus similarly as done above, using the latest PHOENIX model
atmospheres \citep{Husser2013}, 
we explored the possibility of alpha element enhancement 
as the cause of the deep secondary eclipse (i.e., instead of due to a high effective temperature).  
As shown in Fig.~\ref{atms}, 
we tested the upper and lower limit of $\alpha$-element enchancement considered
by the stellar atmospheres, $-0.2$ and $+1.2$ dex, respectively.  
We find that even an enhancement of $+1.2$ dex
is not sufficient to reconcile the observed depth of the secondary eclipse with that of a 
secondary M dwarf with a temperature as expected from theoretical evolutionary models 
\citep[$\sim$3350~K;][]{Baraffe1998,Dotter2008}. 

\item{\bf Contamination by tertiary component or background/foreground star}:
An unresolved component would affect the measured relative depth of the eclipses, 
and thus the derived \tprim/\tsec.  
If there were an unresolved low-mass star, 
then the primary eclipse depth would 
not be severly affected by the third light contamination due to the high luminosity ratio, while 
the secondary eclipse in the NIR would be shallower.  This effect would
cause the measured \tsec\ to be cooler than expected (i.e., not hotter as we observe). 
A more exotic blend, such as that caused by a white dwarf, 
would make primary eclipse shallower in the I-band and 
not significantly affect secondary eclipse depth in the NIR 
causing the secondary star to appear hotter. 
The radial velocity measurements would not be affected in the case of contamination of a background/foreground star.  In the case of a 
physically-associated tertiary component, either a low-mass star or a white dwarf in a long enough period (relative to our RV sensitivity and timespan of our observations) would 
not cause a significant RV motion. 
However, the low-mass component ($\sim$0.14\msun) of the only other EBLM-type object with a measurement of the M-dwarf temperature was  
discovered from its Kepler light curves in which the secondary eclipse was evident 
\citep[KIC 1571511;][]{Ofir2012}. 
This object was also found to be much hotter than expected by the 
DSEP models for its measured mass  ($\Delta \teff \sim$900~K). 
In this case, either blend scenario would dilute both eclipse depths equally because they are observed in the same filter; the temperature ratio for this object would not be affected by either a low-mass star blend or by an exotic blue blend. 
In the case of \eblm, contamination by a low-mass star is not able to explain
the observed hot \tsec, while a blend with a white dwarf
could. 
However, the M-dwarf component of KIC 1571511 is also found to be
significantly hotter than expected, and in that case no blend scenario is able to explain it.  
Thus, we conclude that a blend scenario is not likely the cause of the hot secondary of \eblm;
more measurements of the temperature of M dwarfs would be able to confirm this high 
temperature trend.

\item {\bf Magnetically-induced starspots}:  
Low-mass M dwarfs  
are thought to be very spotted, magnetically-induced spots are expected to be cooler than the photosphere, 
reducing the average temperature while increasing the stellar radius \citep[e.g.,][]{Chabrier2007}.  
Neither of these two effects of starspots would explain the measured, lower-than-expected temperature ratio (i.e., the high secondary temperature), given
that \tprim/\tsec\ is independent from the stellar radii and a cooler effective temperature of 
the secondary would lead to an increase of \tprim/\tsec. 
Although the presence of hot spots (plages) due to activity is not certain on M-dwarfs, several studies have shown that some eclipsing binary light curves are better reproduced with hot spots 
\citep[e.g.,][]{Lopez-Morales2005, Morales2009}. 
However, either a cool spot on the primary component or a bright plage on the secondary would need to be too large to explain the temperature ratio of the system or the secondary eclipse depth. On the contrary, the light curves do not show a significant modulation due to spots. Therefore, we conclude that starspots
are not likely the cause of the lower-than-expected temperature ratio.

\item {\bf Hot spot due to irradiation from the primary star}: It is also not likely that the deeper than expected secondary eclipse is due to a hot spot on the secondary caused by the irradiation from the primary component.  
Because of the eccentric orbit, 
 the components are not rotating synchronously
with the orbital motion, and thus there is not one side of the secondary 
that is always being irradiated by the primary.  In fact, from the measurement
of \vsini\ and the stellar radius, the rotation period of the primary
component is consistent with the pseudo-synchronous rotation ($\sim$12 d; \S \ref{m1}).
However, we have considered the most conservative case in which both components are
rotating synchronously and the mutual irradiation effects between the components are taken into account in the EB modelling.  There is no evidence of a hot spot in the observed and model light curves.
Furthermore, when modeling the light curve with a range in albedo from 0.0 to 1.0 for either star 
the depth of the secondary eclipse remains constant to within 0.001\% in flux.  Thus it is unlikely that a 
hot spot due to irradiation from the primary star can be the cause of
the larger than expected depth of the secondary eclipse.

\item {\bf Residual heat from formation}: It is unlikely that the hot M dwarf temperature we observe is due to residual heat from its formation (i.e., the M dwarf is younger than we estimate).  Firstly, there are no youth signatures in the spectrum of the primary, and close binaries are generally 
thought to be formed at the same time
\citep[e.g.,][]{Prato2003}.  Moreover, M dwarfs of 0.2 \msun\ are thought to stay at roughly a constant temperature from the first several Myrs until the end of their main-sequence life \citep[e.g.,][]{Baraffe2002}, and thus it is never expected during its evolution before and through the main sequence to have such a high temperature.  

\item {\bf Mass transfer and/or accretion}:  
The stellar components of \eblm\ are well detached and inside their Roche lobes.  They
are not transfering mass, and thus are not interacting. There is also no signature in the light curves or
spectra of the presence
of a circumbinary or circumstellar disk from which the stars could be accreting new material. 
Right after formation the radii of the stars could have been large enough to interact and 
they could have
surrounding material from which to accrete.  However, the effects of episodic accretion 
and/or mass transfer at these young ages disappears after a few 
Myrs \citep{Baraffe2009}.  
Thus, given \eblm's old age and that mass transfer and accretion are not currently occuring, 
accretion and/or mass transfer are not likely the cause of the temperature difference.

\item {\bf Tidal heating}: Tidal heating cannot explain the apparent temperature discrepancy.  
Given the orbital period and significant eccentricity, we
examined the possibility that tidal heating could contribute extra
energy to the M dwarf and raise its temperature to the observed
value.  We calculated the present rate of tidal heating with the
``equilibrium tide model,'' which treats the star as a deformed
spheroid and dissipation is determined by just a single parameter,
such as the tidal $Q$ \citep[e.g.][]{Zahn75,FerrazMello2008,Leconte2010}. 
In particular, we use the ``constant-phase-lag'' (CPL) model
\citep{Greenberg2009} as described in \cite{Barnes13}, \citep[see also][]{Heller2011}, and refer
the reader to the former reference for a complete description of this
model. 

In the CPL framework, the rate of dissipation is inversely
proportional to $Q$, which for stars probably has a value in the range
$10^4$--$10^9$
\cite[e.g.][]{MardlingLin02,Jackson09,Matsumura10,Adams11}. Tidal
heating is also a function of the rotation rate, which is unknown for
the M dwarf, and thus we treat it as a free parameter. 

We considered a range of tidal $Q$s from $10^{4}$ and $10^{7}$ and
rotation periods from 1 to 100 days and calculated the tidal heating
rate. The largest tidal heating occurs at the smallest period and $Q$
and is about $4~\times~10^{28}$~ergs/s, about 2.5 orders of magnitude
too low. We also examined non-zero obliquities and found that heating
from obliquity is negligible. We conclude that tidal heating cannot
explain the apparent temperature discrepancy.

\item {\bf Missing physics in atmosphere and/or evolutionary models}:
Although stellar atmosphere and evolutionary models are able to reproduce some of the direct measurements of M dwarfs (e.g., Fig.~\ref{models}), 
as discussed below, the comparison of other mass, radius, metallicity and temperature measurements for M dwarfs are marred
with different assumptions for the distinct kinds of systems (EBLMs, M+M EBs or single M dwarfs). 
There is always the possibility that either or both the atmospheric and evolutionary models
are missing relevant physics for systems such as \eblm\ (e.g., in unequal-mass
binaries, at low metallicities, and/or with a low-mass M dwarf).  A thorough 
assessment of the models is 
beyond the scope of this paper.

\end{itemize}

On the right side of Fig.~\ref{models}, we compare the known measurements for stars 
with masses $<$ 1.0 \msun\ with the BCAH98 and DSEP evolutionary models. 
Typically, these comparisons are done against the  double-lined EBs from which 
masses, radii, and temperatures are measured, and thus, luminosities are derived
\citep[black-filled circles:][]{Bass2012,Birkby2012,Blake2008,Brogaard2011,Carter2011,Coughlin2012,
Creevey2005,Fleming2011,Hartman2011,Hebb2006,Helminiak2011a,Helminiak2011b,
Helminiak2012,Irwin2009,Irwin2011,Kraus2011,Liakos2011,LopezMorales2007,Morales2009,
Rozyczka2009,Thompson2010,Torres2010,Torres2002,Vaccaro2007,Young2006}.
We also include in this comparison the {\it Kepler} EBs with circumbinary planets \citep[green-filled triangles:][]{Doyle2011,Orosz2012,Welsh2012,Schwamb2013}, and KIC 1571511 \citep{Ofir2012} marked
by the yellow diamond which is the only other single-lined EB with measurements 
of the mass, radius {\it and} temperature for the M-dwarf companion (apart from \eblm).
The direct radius measurements from EBs typically find the stars
to be larger than predicted by evolutionary models 
\citep[e.g.,][]{LopezMorales2007,Morales2008,Torres2010}. 
Direct measurements of the radii via interferometry for 
nearby stars are also used for comparison
\citep[brown open squares:][]{Segransan2003,Berger2006,Demory2009,boyajian2012}, and exhibit the same trend. 
It is thought this is due
to magnetic activity and/or reduction of convection efficiency 
\citep[e.g.,][]{Morales2010,Mullan2001}. 
The increase in radius is compensated with a decrease in 
effective temperature to maintain the stellar luminosity.  
The spread in the measurements of radius and temperature as a function of stellar mass in Fig.~\ref{models} is not only due to the metallicity but age becomes important for more massive stars where the evolution is much faster than for the lower mass stars ($\lesssim$ 0.6--0.7 \msun).

In the case of M dwarfs,  their temperatures as a function of mass and metallicity remain uncertain. 
Most M dwarf temperatures are derived from the analysis of spectra line indices and/or broad-band
SED modelling \citep[e.g.,][]{Rojas2012} for nearby, single M dwarfs
with interferometric radii 
(see brown open squares in Fig.~\ref{models}). 
However, for single stars  
it is not possible to obtain a dynamical measurement of their mass, so their mass
estimates rely on comparing observed properties to empirical and/or theoretical scales \citep[e.g.,][]{Rajpurohit2013}. 
In the case of the EBs that exhibit primary and secondary eclipses, like shown in this paper, 
the temperature is typically measured from the temperature ratio (i.e., from the 
relative depth of the eclipses) in
conjunction with a measure of one of the individual temperatures 
(e.g., from spectral type or stellar characterisation of the primary).  
Typically, for double-lined EBs composed of two M dwarfs, the temperature
ratio is measured from the light curve(s) and the integrated temperature
from colour indices or SED models \citep[e.g.,][]{Torres2002}.  
Getting the system's metallicity
is particularly challenging 
because of the two sets of complex spectral features of the two unresolved M dwarfs.  
Although,  it is this double-lined nature
that allows the direct mass determination for these M+M EBs, it hinders the spectroscopic determination
of individual temperatures and metallicity. 
In the case of EBLMs, the method to derive the temperatures is the same as for EBs.
However, it is more precise because the stellar characterisation of our primary star is 
 well-understood in the solar-type regime, and because of the high luminosity contrast between the components 
allows for high-quality (single-lined) spectra  of the primary to be acquired. 
Systems in the EBLM sample, like \eblm, in which the primary is a solar-type star and the secondary is an M dwarf, will provide a large number of measurements of the mass, radius, temperature, metallicity and age for  M dwarfs.

\section{Summary}

In this second paper of the EBLM Project, we derive the orbital parameters of \eblm, 
and the fundamental properties of its stellar components. 
We present the first full analysis of an EBLM in our sample of $\sim$150 systems
discovered from their WASP light curves, thereby defining the project's methodology. 
\eblm\ is an old and metal-poor system, as determined by the large radius and the spectrum
of the solar-type primary star, with an eccentric and long-period orbit. 
The secondary radius of the low-mass M~dwarf is consistent with stellar
evolution models for its given mass, but its temperature is measured to be $\sim$600~K hotter
than expected.  We discard different sources of possible error in our measurement 
of the M~dwarf temperature, including
the treatment of model atmospheres in the eclipsing binary modelling, the stellar
characterisation of the primary, and contamination by an unresolved star.  
We also discuss different physical processes that could have an impact on the M~dwarf affecting
its effective temperature (e.g., hot or cold spots, younger age) or by providing
an additional source of energy (e.g., tidal heating, mass transfer, accretion).  
These scenarios are not able (or are not very likely) to 
account for such a large difference between the temperature expected
by the stellar evolutionary models and the one measured from the secondary eclipse. 
Until the relationship between the mass, radius {\it and} temperature 
for M~dwarfs as a function of their metallicity is well defined, 
 caution must be taken when deriving M~dwarf masses 
 from luminosities, temperatures, and/or colours.  
The EBLM Project will be able to provide these empirical constraints which will 
be crucial, for example, when deriving physical properties of planets around M~dwarfs
discovered by TESS and/or Gaia.

\begin{acknowledgements}
We would like to thank the anonymous referee for providing constructive feedback and comments that 
significantly improved this manuscript. 
The authors would like to thank Isabelle Baraffe, Ignasi Ribas and Barry Smalley for helpful discussions. 
The research leading to these results has received funding from the European Community's Seventh Framework Programme (FP7/2007-2013) under grant agreement number RG226604 (OPTICON).
LHH acknowledges funding support from NSF grant, NSF AST-1009810.
AHMJ Triaud received funding from the Swiss National Science Foundation in the form of an Advanced Mobility post-doctoral fellowship (P300P2-147773).
EGM was supported by the Spanish MINECO project AYA2012-36666 with FEDER support.
RD and SM acknowledge funding support from the Center for Exoplanets and Habitable Worlds. The Center for Exoplanets and Habitable Worlds is supported by the Pennsylvania State University, the Eberly College of Science, and the Pennsylvania Space Grant Consortium. The Hobby--Eberly Telescope (HET) is a joint project of the University of Texas at Austin, the Pennsylvania State University, Stanford University, Ludwig-Maximilians Universitat M\"unchen, and Georg-August-Universit\"at G\"ottingen. The HET is named in honour of its principal benefactors, William P. Hobby and Robert E. Eberly.
This research is based on observations made with the Nordic Optical Telescope, operated by the Nordic Optical Telescope Scientific Association at the Observatorio del Roque de los Muchachos, La Palma, Spain, of the Instituto de Astrofisica de Canarias, as well 
as from Kitt Peak National Observatory, National Optical Astronomy Observatory, which is operated by the Association of Universities for Research in Astronomy (AURA) under cooperative agreement with the National Science Foundation. 
FLAMINGOS was designed and constructed by the IR instrumentation
	group (PI: R. Elston) at the University of Florida, Department of
	Astronomy, with support from NSF grant AST97-31180 and Kitt Peak
	National Observatory.
The BYU West Mountain Observatory 0.91 m telescope was supported by NSF grant AST–0618209 during the time these observations were secured. 
This work was conducted in part using the resources of the Advanced Computing Center for Research and Education at Vanderbilt University, Nashville, TN.
\end{acknowledgements}

\bibliographystyle{aa}
\bibliography{Faedi,extra,leslie}

\end{document}